\documentclass[showpacs,amsmath,amssymb]{revtex4}
\usepackage{graphicx}
\usepackage{epsfig}
\usepackage{color}
\usepackage{graphicx}
\usepackage{dcolumn}
\usepackage{bm}
\begin{document}
\baselineskip=18pt
\large
\title{ The generalized Newton's law of gravitation versus the general theory of relativity}
\author{A. I. Arbab}
\email{aiarbab@uofk.edu} \affiliation{Department of Physics,
Faculty of Science, University of Khartoum, P.O. Box 321, Khartoum
11115, Sudan}
\date{\today}
\begin{abstract} Einstein general theory of relativity (GTR) accounted well for the precession of the perihelion of planets and binary pulsars. While the ordinary Newton law of gravitation failed, a generalized version yields  similar results. We have shown here that these effects can be accounted for as due  to the existence of gravitomagnetism only, and not necessarily due to the curvature of space time. Or alternatively, gravitomagnetism is equivalent to a curved space-time. The precession of the perihelion of planets and binary pulsars may be interpreted as due to the spin of the orbiting planet ($m$) about the Sun ($M$)\,. The spin ($S$) of planets is found to be related to their orbital angular momentum ($L$) by a simple formula, \emph{viz}., $S\propto \,\frac{m}{M}L$\,.
\end{abstract}

 \maketitle
\section{Introduction}
 We have  recently introduced gravitomagnetism as a true cause of the precession of the perihelion of the orbit of planets and binary pulsars [1]. Einstein attributed these effects to the curvature of space-time.  The effect of gravitomagnetism, in a similar manner to electromagnetism, is the Larmor precession of a gravitational moment in the gravitomagnetic field induced by the Sun on the planets.

 Le Verrier discovered that the orbital precession of the planet Mercury was not quite what it should be; the ellipse of its orbit precesses by some minute value than the predicted by the  Newtonian theory of gravitation, even after all the effects of the other planets had been accounted for [1]. This value amounts to  43 arcseconds  per century.  Several classical explanations were put forward, e.g., an interplanetary dust, an unobserved oblateness of the Sun, an undetected moon of Mercury, or a new planet named Vulcan. Others suggested that the Newton inverse-square law is not correct, and accordingly proposed a power law with an exponent that  slightly differs from 2. Moreover, some authors argued in favor of  a velocity-dependent potential (see [1] and references there in).

 To resolve the above mentioned dilemmas, Einstein used a  pseudo-Riemannian geometry to allow for the curvature of space-time that was necessary for the reconciliation of the observed gravitational phenomena. He concluded that the space-time should be curved in order to reproduce the  observed physical laws of gravitation. Owing to Einstein's theory of general relativity, particles of negligible mass travel along geodesics in the space-time.
An exact solution to the Einstein field equations is the Schwarzschild metric, which corresponds to the external gravitational field of a stationary, uncharged, non-rotating, spherically symmetric body of mass $M$. It is characterized by a length scale $r_s$, known as the Schwarzschild radius.  The immediate solutions of the field equations explained the anomalous precession of Mercury, and predicted the observed  bending of light, which were later confirm experimentally [2].

On the other hand, the theory of electromagnetic interaction is accomplished by Maxwell. This is coined  in the four Maxwell equations relating  the electric and magnetic fields to the electric charges and current. Lorentz then obtained the force experienced by a charged particle in electric and  magnetic fields. Larmor has found that when an electron (magnetic moment) is placed in an external magnetic field, the magnetic moment precess about the magnetic field direction. This precession is due to the spin of the electron. This effect is prominent in the spin-orbit interaction exhibited by hydrogen atom [3, 4].

If we now consider gravitation with some scrutiny, we will find that, unlike electromagnetism, moving mass doesn't create a magnetic-like field. Thus, Newton law of gravitation is not like Lorentz law of electromagnetism. In this sense, gravity and electromagnetism are not analogous and can't be utterly compared with gravitation. To remedy this problem, we have to look for a gravitomagnetism counterpart of gravity. In this way, we can say gravity is analogous to electricity and gravitomagnetism is analogous to magnetism. The question is what is the gravitomagnetic field?
By analogy, this should be obtained by looking at Biot-Savart law that defines the magnetic field of a uniformly moving charged particle in an electric field. To complete the analogy the charge of the particle should correspond to the mass of the particle. In this way, we may call the electric charge, the \emph{electric mass} in contrast with the gravitational mass. This furnishes the complete analogy between gravitation and electromagnetism.

How we then avail the electrical phenomena and rules in one paradigm to interpret the other?
To answer this question, we have to trust (beforehand) the existing analogies, and base all our new interpretation of the gravitational phenomena by explaining their corresponding ones. In this manner, the precession of the perihelion of the orbit of planets and binary pulsars is obtained from the precession of the electron (magnetic moment) in an external magnetic field.  Planets and binary pulsars precess when they experience a gravitomagnetic filed (if any). In this case, we use the same laws holding for the counter (analogous) phenomenon, however.

Moreover, the deflection of light by the Sun is explained by using the laws governing the deflection of  a charged particle ($\alpha$-particle) by the  nucleus [5]. If we continue in this manner, we may persuade our selves that, to every electromagnetic phenomena there are gravitomagnetic counter-phenomena. Hence, electromagnetism and gravitomagnetism are same but different aspects of a unified origin.

In this respect, we will find our-selves distracted to interpret the gravitational phenomena as due to the curvature of space-time. We are then not abide by the GTR to interpret our physical world. Or alternatively, we treat the curvature and gravitomagnetism  as a same object, or yield the same effects.
This can be trusted if we are able to show that the  term responsible for the curvature of space-time in Einstein field equations is the same as the that resulting from the influence of gravitomagnetism.

In this paper, we will show that the gravitomagnetism terms in the generalized Newton law of gravitation is the same as the one in the Einstein general field equations. In this way, we upgrade Newton law of gravitation to the general theory of gravitation, but with different  predictions.
Thus, the \emph{correct} Newton law  of gravitation still works finely, and expresses  gravitational phenomena in accordance with observations. Hence, gravity and electromagnetism are governed by unified laws. \textcolor[rgb]{0.00,0.00,1.00}{In section II we present the potential that gives rise to the precession of perihelion in the GTR. We compare this potential with that arising from the gravitomagnetic field. We find that  the gravitomagnetic term is  $\frac{\pi}{3}$ of the Einstein term (GTR). Einstein attributed this term to curvature of space.}

Can we say that the gravitomagnetism is the cause of Einstein curvature?

Do we still adopt GTR, that requires advanced mathematics, as the theory of gravitation and leave the simply-understood Newton's laws of gravitation aside?
\textcolor[rgb]{0.00,0.00,1.00}{In effect, the gravitomagnetic theory (or Gravitational Lorentz force) is simple and can easily be handled with recourse to tensor (advanced mathematics) analysis to unravel gravitational phenomena. Besides, it is analogous to electromagnetic theory that is well understood and comply utterly with experimental facts. The idea of curvature of space is no longer adopted. Moreover, the Einstein's dream of unification of fundamental forces in nature will become imminent within this framework.}
\section{The general theory of relativity (GTR)}
Einstein attributed the gravitational phenomena, now known, to the effect of the curvature of space-time induced by the presence of a massive object [2].
The effective gravitational potential of the object of mass $m$ moving around  a massive object of mass  $M$ takes the form [6]
\begin{equation}
U(r) = -\frac{GMm}{r} + \frac{ L^2 }{ 2 m r^2 } - \frac{GM L^2 }{ c^2 m r^3 }\,,
\end{equation}
and the force, $F=-\frac{\partial U}{\partial r}$\,, can be written as
\begin{equation}
F(r) = -\frac{GMm}{r^2} + \frac{ L^2 }{  m r^3 } - \frac{3GM L^2 }{ m\,c^2 r^4 }\,,
\end{equation}
where $L$ is the orbital angular momentum of the mass $m$.

This inverse-cubic energy term in eq.(1) causes elliptical orbits to precess gradually by an angle $\delta\varphi$ per revolution [2]
\begin{equation}
\delta \varphi = \frac{6\pi G M}{c^2 a \left( 1 - e^{2} \right)} \,,
\end{equation}
where $e$ and $a$ are the eccentricity and semi-major axis of the elliptical orbit, respectively. This is known as the anomalous precession of the planet Mercury.

Another  prediction famously used as evidence for GTR, is the  bending of light in a gravitational field. The deflection angle is given by [2]
\begin{equation}
\delta \theta = \frac{4 G M}{c^2 b} \,,
\end{equation}
where $b$ is the distance of closest approach of light ray to the massive object.
Therefore, the gravitomagnetic force is equal to $\frac{\pi}{3}$ of the GTR force. Whether, the gravitational phenomena are in full agreement with our gravitomagnetic model or with GTR is a subject of the present and future observations. At any rate, we are lucky to have two complementary paradigms explaining the same effect in different ways. Can we deduce that it is the gravitomagnetic field that curves the space and \emph{not} the Sun mass? Or can we say that it is the curvature that produces the gravitomagnetism?
\section{The generalized Newton law of gravitation}
We have shown recently that Newton law of gravitation can be written, as a Lorentz-like law, as [7]
\begin{equation}
\vec{F}(r) =  m\vec{E}_g+m\vec{v}\times \vec{B}_g\,,\qquad E_g=a=\frac{v^2}{r}\,,
\end{equation}
where
\begin{equation}
\vec{B}_g=\frac{\vec{v}\times \vec{E}_g}{c^2}\,.
\end{equation}
Thomas introduced a  factor $\frac{1}{2}$ to account for the spin-orbit interaction in hydrogen atom [8].  Here $B_g$ is measured in $s^{-1}$. To convert it to rad/sec, we multiply it by $2\pi$. Hence, the gravitomagnetic force  becomes
\begin{equation}
F_m(r) = - \frac{\pi\, mv^4}{ c^2 r}\,, \qquad a=\frac{v^2}{r}\,,\qquad v^2=\frac{GM}{r}\,.
\end{equation}
\textcolor[rgb]{0.00,0.00,1.00}{The gravitomagnetic field is divergenceless, since $$\vec{\nabla}\cdot\vec{B}_g=\frac{1}{c^2}\vec{\nabla}\cdot(\vec{v}\times\vec{E}_g)=\frac{1}{c^2}\vec{E}_g\cdot(\vec{\nabla}\times\vec{v})-\frac{1}{c^2}\vec{v}\cdot(\vec{\nabla}\times\vec{E}_g)=-\frac{1}{c^2}\vec{v}\cdot\frac{\partial\vec{B}_g}{\partial t}=-\frac{1}{c^2}\frac{\partial}{\partial t}(\vec{v}\cdot\vec{B}_g)=0\,.$$
This implies that the gravitomagnetic lines curl around the moving mass (gravitational current) creating it. This may also rule out the existence of negative mass. Therefore, as no magnetic monopole exits; no gravitomagnetic monopole (antigravity) exits. Thus, the search for magnetic monopole is tantamount to that of antigravity.}

The angular momentum is defined by $L=mvr$, so that eq.(7) becomes
\begin{equation}
F_m(r)=-\frac{\pi\, GM L^2 }{ m\,c^2 r^4 }\,.
\end{equation}

The second term in eq.(2) is due to the centrifugal term arising from a central force field.
In polar co-ordinates the  force is written as
\begin{equation}
m\vec{a}=m(\ddot r-r\,\dot\theta^2)\, \hat{e}_r+m(r\,\ddot\theta+2\,\dot r\,\dot\theta)\,\hat{e}_\theta\,.
\end{equation}
For a central force the second term vanishes. It yields, $\dot\theta=\frac{L}{mr^2}$\,, so that the first term becomes
\begin{equation}
m\,a_r=m\,\ddot r-\frac{L^2}{m\,r^3}\,.
\end{equation}
Substituting eq.(10) in eq.(5) yields  the full effective central force, owing to gravitomagnetism, as
\begin{equation}
F(r) = -\frac{GMm}{r^2}+\frac{L^2 }{m r^3} -\frac{\pi\, GM L^2}{ m\,c^2 r^4}\,.
\end{equation}
The corresponding potential will be
$$
U(r) = -\frac{GMm}{r}+\frac{L^2 }{2m r^2} -\frac{\pi\, GM L^2}{3\,\, m\,c^2 r^3}\,.
$$
Comparison of eqs.(2) and (11) reveals that  the gravitomagnetic force is equal to $\frac{\pi}{3}$ of the curvature force. Consequently, the generalized Newton law of gravitation and the general theory of relativity produce the same gravitational phenomena.

The gravitomagnetic force term, the last term in eq.(11), can be written as
\begin{equation}
\frac{\pi\, GM L^2}{ m\,c^2 r^4}=\frac{\pi\, G^2M^2m }{c^2 r^3}\,\,,\qquad {\rm where,}\,\,\,v^2=\frac{GM}{r}\,.
\end{equation}
Finally, eq.(11)  can be written as
\begin{equation}
F(r) =- \frac{GMm}{r^2}+ \frac{ J_{\rm eff.}^2 }{m r^3} \,,
\end{equation}
where
\begin{equation}
 J_{\rm eff.}^2=L^2-\left(\frac{\sqrt{\pi}GMm }{ c} \right)^2\,.
\end{equation}

\section{Precession of planets and binary pulsars}
Owing to the above equivalence between gravitomagnetism and GTR, we interpret the precession of the perihelion of planets and binary pulsars as a Larmor-like precession, and not due to the GTR interpretation as due to the curvature of space-time. We may attributed this precession as due to the precession of  gravitational moment (mass) in a gravitomagnetic field induced by the massive objects (Sun). In electromagnetism, the Larmor precession is defined by [4]
\begin{equation}
\omega=\frac{e}{2m}B\,,
\end{equation}
while  in gravitation (since $B_g$ is in $s^{-1}$ and $e\Leftrightarrow m$) it is defined as [1]
\begin{equation}
\omega_g=2\pi\left(\frac{B_g}{2}\right)=\frac{\pi v^3}{rc^2} \,, \qquad B_g=\frac{v\,a}{c^2}=\frac{v^3}{rc^2}\,,
\end{equation}
where ($\omega_g$ is in rad/seec) and
\begin{equation}
a=\frac{v^2}{r}\,.
\end{equation}
The precession rate in eq.(16) can be written as
\begin{equation}
\omega_g=\pi\left(\frac{2\pi\, GM}{T\,c^2r}\right)=\frac{\delta\varphi_g}{T}\,,
\end{equation}
where $T=\frac{2\pi r}{v}$ is the period of revolution. This corresponds to a precession angle of
\begin{equation}
\delta\varphi_g=3\left(\frac{2\pi\,GM}{c^2r}\right)\, \rm rad/s\,,
\end{equation}
that is equal to $\frac{\pi}{3}$ of the curvature effect, and for elliptical orbit  $r=a(1-e^2)$.
\section{Deflection of $\alpha$-particles by the nucleus}
We would like here to interpret the deflection of light by the Sun gravity in an analogous way to the deflection of $\alpha$-particles by the nucleus, without resorting to the GTR calculation. The deflection angle of $\alpha$-particles by a nucleus is given by [5]
\begin{equation}
\Delta\theta_e=\frac{4keQ}{mb\,v^2}\,,
\end{equation}
where $Q$ is the nucleus charge, $v$ the $\alpha$-particle speed, $k$  Coulomb constant,  and $b$ the impact factor. The corresponding gravitational analog for the deflection of light will be, $v\rightarrow c$, $e\rightarrow m$, $Q\rightarrow M$, $k\rightarrow G$, [9]
\begin{equation}
\Delta\theta_g=\frac{4GM}{b\,c^2}\,,
\end{equation}
without resorting to GTR calculation.  Recall that, according to Equivalence Principle, all particles in gravity accelerate without reference to their mass (whether massive or massless). Therefore, it doesn't matter wether light has a mass or not.
The relation in eq.(21) is the same as the relation obtained by GTR as in eq.(4). The minimum distance$\alpha$ particles can approach the nucleus is given by equating the kinetic energy and the Coulomb potential energy that yields the relation
\begin{equation}
b_e=\frac{2kq_1q_2}{mv^2}\,.
\end{equation}
In gravitation and for light scattered by the Sun gravity, the above relation gives ($q_1\rightarrow m$, $q_2\rightarrow M$ and $k\rightarrow G$)
\begin{equation}
b_g=\frac{2GM}{c^2}\,.
\end{equation}
This is nothing but the Schwarzschild distance that no particle can exceed.
Therefore, the complete analogy between gravitation and electricity is thus realized. In this context, we have shown recently that the Larmor dipole radiation has a gravitational analogue [10]. Similarly, the same analogy exists between hydrodynamics and electromagnetism [11].

\section{The spin of planets}
\textcolor[rgb]{0.00,0.00,1.00}{The discovery of the spin of the electron by  Goudsmit and  Uhlenbeck  in 1926 was crucial in understanding many physical phenomena that wouldn't have been explained without [12]. This spin is theoretically formulated by Dirac confirming the experimental finding . However, the spin of planets had been known since long time (1851) that was demonstrated  by Foucault's pendulum. In a recent paper we have introduced the gravitomagnetism produced by moving planets as the magnetic field produced by moving charge [1]. We then obtained the gravitational Ampere's and Faraday's laws of gravitomagnetism. The gravitomagnetic moment of a planet due to its orbital motion is given by [1]}
\begin{equation}
\mu_{L}=\frac{v^3r^2}{2G}\,.
\end{equation}
For circular orbit, eq.(24) yields
\begin{equation}
\mu_{L}=\left(\frac{M}{2m}\right)\, L\,.
\end{equation}
\textcolor[rgb]{0.00,0.00,1.00}{In a similar manner the gravitomagnetic moment due to spin will be twice the above value (analogous to electromagnetism)}
\begin{equation}
\mu_{S}=g_S\left(\frac{M}{2m}\right)\, S\,,
\end{equation}
\textcolor[rgb]{0.00,0.00,1.00}{where $g_S$ defines some gyro-gravitomagnetic ratio that is independent of the planet's mass. If we assume the precession of planets is an spin-orbit interaction, then we can equate $-\mu_SB_g$ (assuming the angle to be zero) to the potential term arising from the gravitomagnetic force in eq.(11). This yields, for circular orbit,}
\begin{equation}
S=\left(\frac{4\pi}{3g_S}\frac{m}{M}\right)\, L\,\,,\qquad \qquad S=\left(\frac{4\pi}{3g_S}\frac{Gm^2}{v}\right).
\end{equation}
\textcolor[rgb]{0.00,0.00,1.00}{This is a very interesting equation, since it determines the spin of planets from their orbital angular momentum. With the help of the above equation, the moment of inertia of planets can be precisely determined. It then follows that the spin and the geometrical form of  planets is a consequence of its dynamics. Consequently, the spin angular momentum is no longer an intrinsic property of the planet. The energy corresponding to  this interaction may be converted into internal energy (heat) inside the planet. }

\textcolor[rgb]{0.00,0.00,1.00}{Owing to eq.(27) we are entitled to say that any orbiting planet must spin! Thus, any gravitating object in curvilinear motion must spin. For consistency of the spin of the Earth with the present value with take $g_S=57$.  From this law the moment of inertia of all gravitation objects can be  precisely determined. Table 1 shows the anticipated values for the spin and the corresponding moment of inertia of the planetary system. Equation (27) can be used to estimate the hidden central mass around which another mass orbits. It can be generally useful in many astrophysical applications.}

\begin{table}
\begin{tabular}{|l|l|c|}
  \hline
  Planet & Spin ($\rm Js$) & Moment of inertia $(\rm Kgm^2)$ \\
\hline \hline
Mercury & 1.12 E+31 &  8.98 E+36\\
Venus & 3.31 E+33 & 1.10 E+40 \\
Earth & 5.84 E+33 & 8.02 E+37 \\
Mars & 8.32 E+31 & 1.17 E+36\\
Jupiter & 1.35 E+39 & 7.69 E+42 \\
Saturn & 1.64 E+38 & 1.00 E+42 \\
Uranus & 5.44 E+36 & 5.43 E+40 \\
Neptune & 9.45 E+36 & 8.12 E+40\\
\hline
\end{tabular}
\caption{The predicated values for spin and moment of inertia owing to eq.(27) with  $g_S=57$. Any deviation from known values that may appear could be attributed to the uncertainty in determining the radii of planets. Alternatively,  the angle between $L$ and $S$ will be of importance.}
\end{table}
\section{Conclusions}
We have shown that the gravitomagnetism and the general theory of relativity are two theories of the same phenomenon. This entitles us to fully accept the analogy existing between electromerism and gravity. Hence, electromagnetism and  gravity are unified phenomena. The precession of the perihelion of planets and binary pulsars may be interpreted as a spin-orbit interaction of gravitating objects. The spin of a planet is directly proportional to its orbital angular momentum and  mass weighted by the Sun's mass. Alternatively, the spin is directly proportional to the square of the orbiting planet's mass and inversely proportional to its velocity.
\section*{Acknowledgements}
\textcolor[rgb]{0.00,0.00,1.00}{I would like to thank the anonymous referee for his (her) useful and critical comments. I also thank the referee for drawing my attention to consider a formula for the spin of planets.}

\end{document}